%% file: proc_J.tex
\documentclass[11pt]{article}
\usepackage{cite}
\usepackage{graphics}
\usepackage{pstricks}
\usepackage{pst-node}
\usepackage{here}
\usepackage{axodraw}
\usepackage{epsfig}
\usepackage{epsf}
\usepackage{graphicx}
\usepackage{amssymb}
\def\sig{\left[\frac{\displaystyle{\mathrm{d}\sigma}}{\displaystyle{\mathrm{d}\cos \, \theta}}\right]}
\def \ni {\noindent}
\def \be {\begin{equation}}
\def \e {\end{equation}}
\def \bea {\begin{eqnarray}}
\def \ea {\end{eqnarray}}
\def \eps {\epsilon}

\newcommand{\To}[2]{\stackrel{#1}{\hbox to #2 pt{\rightarrowfill}}}

\def\wp{\ifmmode W^+\else $W^+$\fi}
\def\wm{\ifmmode W^-\else $W^-$\fi}
\def\emm{\ifmmode e^-\else $e^-$\fi}
\def\ep{\ifmmode e^+\else $e^+$\fi}

\def \e#1#2{\ensuremath { \eps_{\,#1}^{\,*}\,({#2})\,\, }}

\def\sig{\left[\frac{\displaystyle{\mathrm{d}\sigma}}{\displaystyle{\mathrm{d}\cos \, \theta}}\right]}

\newcommand{\bqa}{\begin{eqnarray}}
\newcommand{\eqa}{\end{eqnarray}}
\newcommand{\ba}{\begin{eqnarray}}

\newcommand{\sla}[1]{#1 \!\!\! \slash}
\begin{document}
\def\theequation{\arabic{section}.\arabic{equation}}

\vspace{-2cm}
\renewcommand{\thefootnote}{\fnsymbol{footnote}}  
% \begin{flushright}  
\ni
DESY 02-025           \hfill  hep-ph/0203220
\\
KEK-CP-123  \hfill   ISSN 0418--9833
\\  
BI--TP 2002/03 
\\
LMU-01/02
\\
March 2002 
\\
% \end{flushright}  
\vspace{1cm} 
\begin{center}  
{\LARGE \bf One-loop corrections 
% \\[3mm]
to the process \boldmath{$ e^+e^- \to t{\bar t}$} including hard
bremsstrahlung
\footnote{
Talk presented by J. Fleischer.
}
\footnote{
Work supported in part by the European Community's Human Potential
Programme under contract HPRN-CT-2000-00149 Physics at Colliders.
% \\[-1mm]
} 
}
  \\ 
%Work supported by the European Union under
%  contract HPRN-CT-2000-00149 \\[-1mm]}  \\ 
\vspace{1.5cm} 
{ 
%{\large A. Leike}${}^{1}$\footnote{leike@theorie.physik.uni-muenchen.de}, 
%----------------------------------------------------------------------
{\Large J. Fleischer}${}^{1}$\footnote{E-mails:~
fl@physik.uni-bielefeld.de,
Junpei.FUJIMOTO@th.u-psud.fr,
tadashi.ishikawa@kek.jp,
\\
\phantom{E-mails:~~~~~~~}
leike@theorie.physik.uni-muenchen.de,
Tord.Riemann@desy.de,
shimiz@suchi.kek.jp,
\\
\phantom{E-mails:~~~~~~~~}Anja.Werthenbach@desy.de}
~~{\Large J. Fujimoto}${}^{2}$,
~~{\Large T. Ishikawa}${}^{2}$,
~~{\Large A. Leike}${}^{3}$,
~~{\Large T. Riemann}${}^{4}$,
\\
\vspace{0.2cm} 
~~{\Large Y. Shimizu}${}^{2}$ and
~~{\Large A. Werthenbach}${}^{4}$ } 
\\
\vspace{0.6cm} 

%\phantom{E-mails:~~~}~~~Anja.Werthenbach@desy.de

\small 
%{\sf 1)  Ludwig-Maximilians-University Munich, Theresienstr. 37
%    D-80333 Munich, Germany\\ }
{
${}^{1}$~Fakult\"at f. Physik, Universit\"at Bielefeld,  33615
Bielefeld, Germany\\ }
\smallskip
{
${}^{2}$~High Energy Accelerator Research Org.~(KEK),
Oho 1-1, Tsukuba, Ibaraki 305-0801, Japan   \\ }
\smallskip
{
${}^{3}$~%
Sektion Physik der Universit\"at M\"unchen, 80333 M\"unchen, Germany
\\ }
\smallskip
{
${}^{4}$~Deutsches Elektronen-Synchrotron, DESY Zeuthen,  15738 Zeuthen, Germany  }  

%endofsmall
%%%
\vspace{1cm}

~\\
\vspace{1cm} 

{

%{
%${}^{4}$~Department of Physics, Nagoya University, Chikusaku,
%Nagoya 464, Japan   \\ }
%\smallskip
%{
%${}^{6}$~National Laboratory for High Energy Physics, Tsukuba,
%Ibaraki 305, Japan  }  

}%endofsmall

%%%
\end{center}

\vspace{1cm} 

\begin{center}
{\Large \bf \sc{Abstract}}
\end{center}
    Radiative corrections to the process $e^+ e^- \to t{\bar t}$ are
calculated in one-loop approximation of the Standard Model. There exist
results from several groups 
\cite{Fujimoto:1988hu},
\cite{Beenakker:1991ca,Hollik:1998md},
\cite{Andonov:2002rr,Bardin:2000kn},
\cite{FRW:2002sw}.
This talk provides further comparisons of the complete
elektroweak contributions, including hard bremsstrahlung. 
The excellent final agreement of the different groups
allows to continue by working on a code for an event generator 
for TESLA and an extension to $e^+ e^- \to 6$ fermions.
\\

\setcounter{footnote}{0}  
\renewcommand{\thefootnote}{\arabic{footnote}}  
%\vfill  
%\clearpage  
\setcounter{page}{1}  
\thispagestyle{empty} 
%\vspace*{-0.5cm} 

\newpage

%{\Large \bf \sc{Introduction}}

   The process $e^+ e^- \to t{\bar t}$ is one of the most prominent and 
important processes to be measured at TESLA in order to find possible
deviations from the Standard Model and thus `New Physics'. Several 
groups have had results for the one-loop corrections, but comparisons 
were not performed in all details. In order to provide finally an event 
generator for the evaluation of experimental data, such comparisons
are mandatory. The {\tt topfit} collaboration \cite{FRW:2002sw} having 
compared before the QED and weak corrections including soft photon
contributions with 
\cite{Beenakker:1991ca,Hollik:1998md}, see Ref. \cite{Fleischer:2002rn}, 
we are now also comparing the hard bremsstrahlung.
This calculation has been performed before by the  GRACE 
group \cite{Yuasa:1999rg} as well 
and in the present comparison we find
excellent agreement between the results of our two groups,
which means that the technical precision of the one-loop
approach is completely tested.
Thus we have 
been able to collect all the necessary codes to go on to more
elaborate calculations, like e.g. inclusion of higher order
corrections and a pole approximation of the
process $e^+ e^- \to 6$ fermions, see e.g. \cite{Aeppli:1994rs,
Kolodziej:2001xe}.\\  

We define the following four-momenta and invariants:
%\begin{center}
$k_{1,2}$ for $e^{-,+}$,
$p_{1,2}$ for ${t},\bar{t}$,
$p$ for the photon $\gamma$,
c.m. energy: $s=(k_1+k_2)^2$, 
$s':=(p_1+p_2)^2$ invariant mass of the top pair and $t=(k_1-p_1)^2$.\\
%\end{center}

In Refs. \cite{Beenakker:1991ca,Hollik:1998md},
the following basis was introduced for the
decomposition of the scattering matrix element into amplitudes:
\bqa
{M_1}^{\rho \kappa} &=& 
\left[\bar{u_t}(p_1) {\gamma}^{\mu} {\omega}_{\rho} v_t(p_2)\right]
\left[\bar{v_e}(k_2) {\gamma}_{\mu} {\omega}_{\kappa} u_e(k_1)\right]
\nonumber\\
{M_2}^{\rho \kappa} &=& 
\left[\bar{u_t}(p_1) {\sla{k}}_2 {\omega}_{\rho} v_t(p_2)\right]
\left[\bar{v_e}(k_2) {\sla{p}}_1 {\omega}_{\kappa} u_e(k_1)\right]
\nonumber\\
{M_3}^{\rho \kappa} &=&
\left[\bar{u_t}(p_1) {\omega}_{\rho} v_t(p_2)\right]
\left[\bar{v_e}(k_2) {\sla{p}}_1 {\omega}_{\kappa} u_e(k_1)\right]
\nonumber\\
{M_4}^{\rho \kappa} &=& 
\left[\bar{u_t}(p_1) {\gamma}^{\mu} {\sla{k}}_2 {\omega}_{\rho} v_t(p_2)\right]
\left[\bar{v_e}(k_2) {\gamma}_{\mu} {\omega}_{\kappa} u_e(k_1)\right],\nonumber
\eqa
where
\bqa
{\omega}_{\lambda} &=& \frac{1+\lambda {\gamma}_5}{2}, \lambda=\pm 1. 
\nonumber
\eqa
   For $m_e=0$ (as assumed here),
only 6 independent amplitudes of the 16 introduced above
remain: there exist relations like
\begin{eqnarray}
{M_4}^{-~+} &=& 
 -m_t \left({M_1}^{+~+}+{M_1}^{-~+}\right) + 2 {M_3}^{-~+}
\nonumber\\
{M_2}^{+~+} &=&
-m_t {M_3}^{-~+} + \frac{1}{2} \left(m_t^2 {M_1}^{-~+} + t {M_1}^{+~+}\right)
\nonumber\\
{M_3}^{+~+} &=&
{M_3}^{-~+}. 
\nonumber
\end{eqnarray}

%The cross-section reads: \\
%\vspace{-1.cm}
%\begin{eqnarray}
%d\sigma(2\to 3) &=&
%\frac{(2\pi)^4}{2 \sqrt{\lambda(s,m_e^2,m_e^2)}} \times \left|{\cal
%  M}\right|^2  
%\nl &&
%\times~ d\Phi_3(p_++p_-;p_1,\ldots,p_3),\nonumber
%\\
%d\Phi_3(p_1,\ldots,p_3)
%&=& 
%\delta^4\left(P-\sum_ip_i\right) \prod_{i=1}^3
%d^4p_i\delta\left(p_i^2-m_i^2\right)\theta(p_i^0),\nonumber
%\\ 
%\sqrt{\lambda(s,m_e^2,m_e^2)} &=& 2\sqrt{(p_+p_-)^2-m_e^4} \approx 2s,
%\nonumber
%\end{eqnarray}
%with the Kallen function:
%\bqa
%\lambda(a,b,c) &=& a^2 + b^2 + c^2 -2ab-2bc-2ca 
%\nonumber
%\eqa
%\newpage
%\begin{center}
%\Huge{Bremsstrahlung}
%\end{center}

One crucial subject of our present comparison is the hard 
bremsstrahlung. Therefore we give an extensive account of 
the bremsstrahlung in general.
The initial- and final-state bremsstrahlung are shown in Figs. 1,2.
%\bigskip
%%%%%%%%%%%%%   All figures for  interference   %%%%%%%%%%%%%
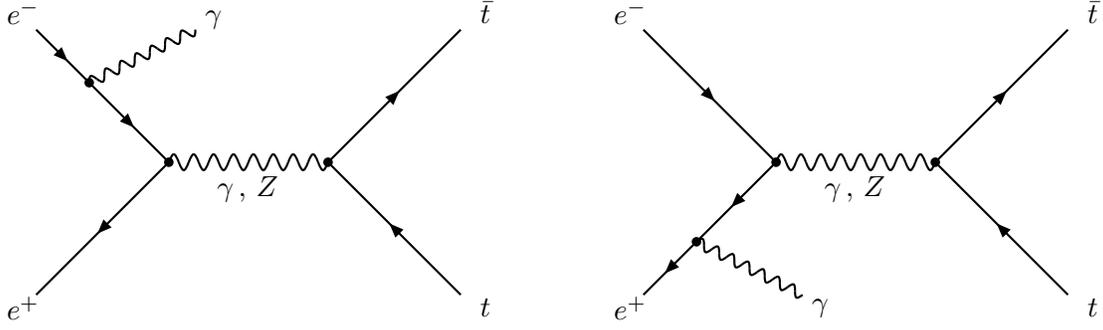
\begin{figure}[htb]
%%%%%%%%%%%%%%%%%%%%%%%%%%%%%%%%%%%
% initial-state radiation diagrams %
%%%%%%%%%%%%%%%%%%%%%%%%%%%%%%%%%%%
%\hspace{-2.0cm}
\begin{minipage}[bht]{7.9cm}
{\begin{center}
%-------------------
\vfill
\setlength{\unitlength}{1pt}
%\SetScale{0.8}
\SetWidth{0.8}
\begin{picture}(180,120)(0,0)
\thicklines
\ArrowLine(10,110)(30,90)
\Vertex(30,90){1.8}
\Photon(30,90)(70,110){2}{7}
\ArrowLine(30,90)(60,60)
\Vertex(60,60){1.8}
\ArrowLine(60,60)(10,10)
\Photon(60,60)(120,60){3}{8}
\Vertex(120,60){1.8}
\ArrowLine(120,60)(170,110)
\ArrowLine(170,10)(120,60)
\Text(5,117)[]{$e^-$}
\Text(5,5)[]{$e^+$}
\Text(90,50)[]{$\gamma\,,\,Z$}
\Text(180,5)[]{$t$}
\Text(180,117)[]{$\bar{t}$}
\Text(77,114)[]{$\gamma$}
\end{picture}
%\thispagestyle{empty}
%----------------
\end{center}}
\end{minipage}
\hspace{0.0cm}
\begin{minipage}[bht]{7.9cm}
{\begin{center}
%-------------------
\vfill
\setlength{\unitlength}{1pt}
%\SetScale{0.8}
\SetWidth{0.8}
\begin{picture}(180,120)(0,0)
\thicklines
\ArrowLine(10,110)(60,60)
\Vertex(60,60){1.8}
\ArrowLine(60,60)(30,30)
\Vertex(30,30){1.8}
\Photon(30,30)(70,10){2}{7}
\ArrowLine(30,30)(10,10)
\Photon(60,60)(120,60){3}{8}
\Vertex(120,60){1.8}
\ArrowLine(120,60)(170,110)
\ArrowLine(170,10)(120,60)
\Text(5,117)[]{$e^-$}
\Text(5,5)[]{$e^+$}
\Text(90,50)[]{$\gamma\,,\,Z$}
\Text(180,5)[]{$t$}
\Text(180,117)[]{$\bar{t}$}
\Text(77,5)[]{$\gamma$}
\end{picture}
\thispagestyle{empty}
%----------------
\end{center}} 
\end{minipage}
%
%\vspace{2.0cm}
\caption[]{\sf\label{Fig.ini} Feynman diagrams for initial-state
radiation }
\end{figure}
%%%%%%%%%%%%%%%%%%%%%%%%%%%%%%%%%%%%%%%%%%%%%%%%%%%%%%%%%%%%%%%%%%%
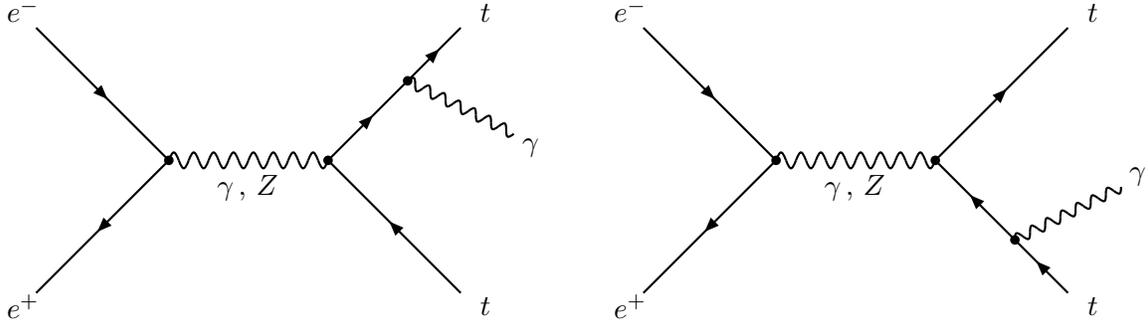
\begin{figure}[htb]
%%%%%%%%%%%%%%%%%%%%%%%%%%%%%%%%%%%
% final-state radiation diagrams %
%%%%%%%%%%%%%%%%%%%%%%%%%%%%%%%%%%%
%\hspace{-2.0cm}
\begin{minipage}[bht]{7.9cm}
{\begin{center}
%-------------------
\vfill
\setlength{\unitlength}{1pt}
%\SetScale{0.8}
\SetWidth{0.8}
\begin{picture}(180,120)(0,0)
\thicklines
\ArrowLine(10,110)(60,60)
\Vertex(60,60){1.8}
\ArrowLine(60,60)(10,10)
\Photon(60,60)(120,60){3}{8}
\Vertex(120,60){1.8}
\ArrowLine(120,60)(150,90)
\Vertex(150,90){1.8}
\Photon(150,90)(190,70){2}{7}
\ArrowLine(150,90)(170,110)
\ArrowLine(170,10)(120,60)
\Text(5,117)[]{$e^-$}
\Text(5,5)[]{$e^+$}
\Text(90,50)[]{$\gamma\,,\,Z$}
\Text(180,5)[]{$t$}
\Text(180,117)[]{$\bar{t}$}
\Text(197,65)[]{$\gamma$}
\end{picture}
%\thispagestyle{empty}
%----------------
\end{center}}
\end{minipage}
\hspace{0.0cm}
\begin{minipage}[bht]{7.9cm}
{\begin{center}
%-------------------
\vfill
\setlength{\unitlength}{1pt}
%\SetScale{0.8}
\SetWidth{0.8}
\begin{picture}(180,120)(0,0)
\thicklines
\ArrowLine(10,110)(60,60)
\Vertex(60,60){1.8}
\ArrowLine(60,60)(10,10)
\Photon(60,60)(120,60){3}{8}
\Vertex(120,60){1.8}
\ArrowLine(120,60)(170,110)
\ArrowLine(170,10)(150,30)
\Vertex(150,30){1.8}
\Photon(150,30)(190,50){2}{7}
\ArrowLine(150,30)(120,60)
\Text(5,117)[]{$e^-$}
\Text(5,5)[]{$e^+$}
\Text(90,50)[]{$\gamma\,,\,Z$}
\Text(180,5)[]{$t$}
\Text(180,117)[]{$\bar{t}$}
\Text(197,54)[]{$\gamma$}
\end{picture}
%\thispagestyle{empty}
%----------------
\end{center}} 
\end{minipage}
%
%\vspace{2.0cm}
\caption[]{\sf\label{Fig.fin} Feynman diagrams for final-state
radiation }
\end{figure}
%------------------------------------------------------------
%\newpage
%\Large{Invariants: \\
Special, often used invariants, are the following:
$Z_1$ and $Z_2$ from the propagators of the 
electron and positron in the diagrams 
of the initial-state radiation,
\begin{eqnarray}
\label{Z1}
 Z_1 &=& 2 p\cdot k_1 = - \left[ (k_1 - p)^2 -  m_e^2 \right]
\nonumber \\
 Z_2 &=& 2 p\cdot k_2 = - \left[ (p - k_2)^2 -  m_e^2 \right]
\nonumber
\end{eqnarray}
and similarly $V_1$ and $V_2$ from the propagators
of the final top and anti-top 
of the final-state radiation,
\begin{eqnarray}
 V_1 &=& 2 p\cdot p_1 = \left[ (p + p_1)^2 -  m_t^2 \right]
\nonumber \\
 V_2 &=& 2 p\cdot p_2 = \left[ (p + p_2)^2 -  m_t^2 \right].
\nonumber
\end{eqnarray}
%\newpage

The differential cross section of the bremsstrahlung reads:\\
\begin{eqnarray}
 d \sigma = \frac{(2 \pi)^4}{2 s} \mid
 M_{ini} + M_{fin} \mid^2 d{\Phi_3},\nonumber
\end{eqnarray}
where $d{\Phi_3} $ is the differential 
phase space of three outgoing particles (see below).
%;$\beta_0=\sqrt(1-\frac{4 me^2}{s}).$\\ 

Of the one-loop vertex corrections we show the QED ones in Fig.~3.
The mechanism of cancellation of the infrared (IR) soft photon 
divergences is as follows:\\

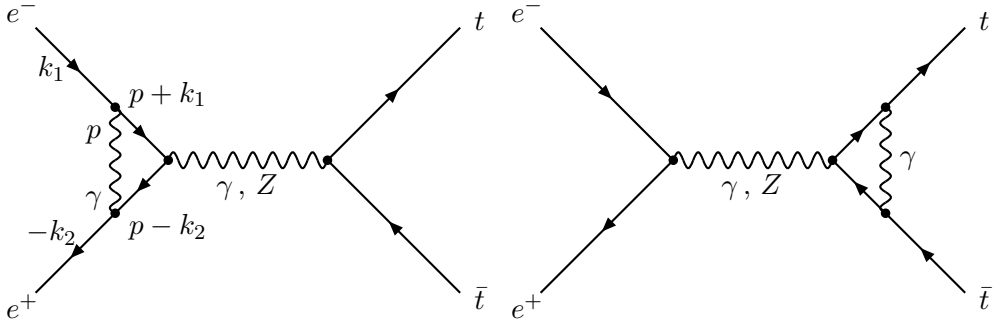
\begin{figure}[tbhp]
\begin{center}
% new vertices:
%%%%%%%%%%%%%%%%%%%%%%%%%%%%%
% Vertex correction diagram zvert1.tex%
%%%%%%%%%%%%%%%%%%%%%%%%%%%%%
\vfill
\setlength{\unitlength}{1pt}
%\SetScale{0.8}
\SetWidth{0.8}
\begin{picture}(180,120)(0,0)
\thicklines
\ArrowLine(10,110)(40,80)
\Vertex(40,80){1.8}
\ArrowLine(40,80)(60,60)
\Photon(40,80)(40,40){2}{5}
\ArrowLine(60,60)(40,40)
\Vertex(40,40){1.8}
\ArrowLine(40,40)(10,10)
\Vertex(60,60){1.8}
\Photon(60,60)(120,60){3}{8}
\Vertex(120,60){1.8}
\ArrowLine(120,60)(170,110)
\ArrowLine(170,10)(120,60)
\Text(5,117)[]{$e^-$}
\Text(5,5)[]{$e^+$}
\Text(90,50)[]{$\gamma\,,\,Z$}
\Text(178,8)[]{$\bar{t}$}
\Text(178,113)[]{$t$}
\Text(32,45)[]{$\gamma$}
\Text(32,70)[]{$p$}
\Text(16,33)[]{$-k_2$}
\Text(16,95)[]{$k_1$}
\Text(60,35)[]{$p-k_2$}
\Text(60,85)[]{$p+k_1$}
\end{picture}
%%%%%%%%%%%%%%%%%%%%%%%%%%%%%
% Vertex correction diagram zvert2.tex%
%%%%%%%%%%%%%%%%%%%%%%%%%%%%%
\setlength{\unitlength}{1pt}
%\SetScale{0.8}
\SetWidth{0.8}
\begin{picture}(180,120)(0,0)
\thicklines
\ArrowLine(10,110)(60,60)
\Vertex(60,60){1.8}
\ArrowLine(60,60)(10,10)
\Photon(60,60)(120,60){3}{8}
\Vertex(120,60){1.8}
\ArrowLine(170,10)(140,40)
\Vertex(140,40){1.8}
\ArrowLine(140,40)(120,60)
\Photon(140,40)(140,80){2}{5}
\ArrowLine(120,60)(140,80)
\Vertex(140,80){1.8}
\ArrowLine(140,80)(170,110)
\Text(5,117)[]{$e^-$}
\Text(5,5)[]{$e^+$}
\Text(90,50)[]{$\gamma\,,\,Z$}
\Text(178,8)[]{$\bar{t}$}
\Text(178,113)[]{$t$}
\Text(149,60)[]{$\gamma$}
\end{picture}

\vspace*{0.5cm}
\caption[Photonic vertex corrections]
{\sf
The photonic vertex corrections. % for initial (a) and final (b) states
}
\label{Fig.virtph}
\end{center}
\end{figure}
%------------------------------------------------------------
%\begin{center}
the interference between the Born and the initial state vertex
diagram cancels the IR-divergences of 
\bqa
\mid M_{ini} \mid^2,\nonumber
\eqa
%\bigskip
the interference between the Born and final state vertex diagram 
cancels the IR-divergences
of 
\bqa
\mid M_{fin} \mid^2, \nonumber
\eqa
%\bigskip
the interference between $M_{ini}$ and $M_{fin}$ is cancelled by box-diagrams
($\gamma \gamma$ and $\gamma Z$ in the s-channel).\\
%\end{center}

%\newpage

\begin{figure}[htp]
\hspace{-4.0cm}
\begin{minipage}[bht]{7.9cm}
{\begin{center}
%-------------------
  \vspace{-3.0cm}
  \hspace{-2.0cm}
  \mbox{
  \epsfysize=16cm
  \epsffile[0 0 500 500]{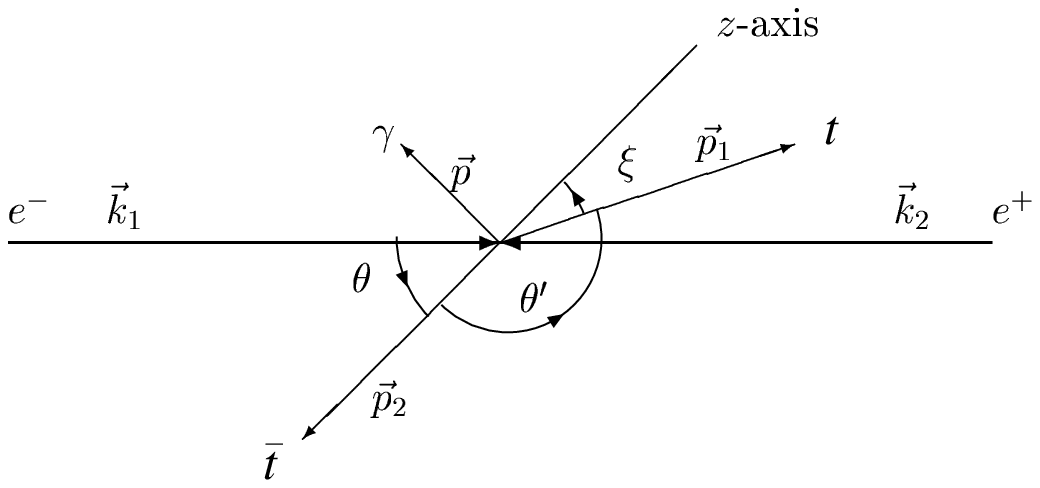}
  }
%----------------
\end{center}}
\end{minipage}
\hspace{2.0cm}
\begin{minipage}[bht]{7.9cm}
{\begin{center}
%-------------------
  \vspace{8.0cm}
  \hspace{0.0cm}
  \mbox{
  \epsfysize=16cm
  \epsffile[0 0 500 500]{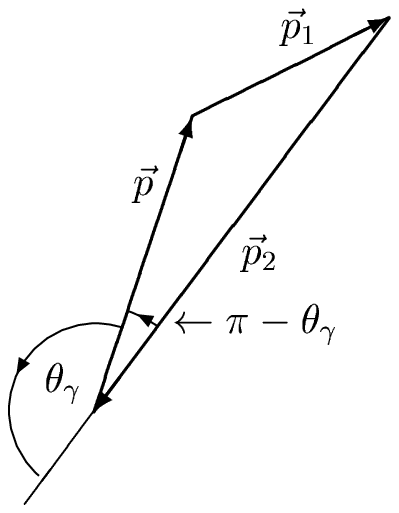}
  }
%----------------
\end{center}} 
\end{minipage}
\vspace{-18.0cm}
\caption[Angles of phase space]
{\label{phfig2}
\sf Angles of phase space and photon angle
  $\theta_\gamma$.} 
\end{figure}

The phase space of the process under consideration may be characterized
by the following four independent kinematical variables:
\begin{itemize}
\label{indvar}
\item $s':=(p_1+p_2)^2$ as invariant mass squared of the top pair,

\item $\cos\theta$ as cosine of the scattering angle of $\bar{t}$ 
      with respect to the $e^-$ beam axis in the c.m.s.~.

\item $\cos\theta_\gamma$ as cosine of the polar angle 
      between the three-momenta of $\bar{t}$ and 
      the photon in the c.m.s., which is 
      related to $V_2$ via an algebraic relation. 

\item $\varphi_\gamma$ as azimuthal angle of the photon in the rest frame of 
      $({t},\gamma)$\\ 
      ($z$-axis defined by $\vec p_2$ of $\bar{t}$ in the
      c.m.s.). 
\end{itemize}
The different angles of the phase space are shown in Fig.~\ref{phfig2}.

Here $\xi$ is the `acollinearity' angle of the final fermions.
$\xi = \pi$ means no
constraint by collinearity, the requirement $\xi \ll 1$ restricts
the kinematics to born-like events: only soft photons or photons
collinear to one of the final fermions are allowed.\\

The integration over the phase space is reduced to an integration
over the above four variables in the following manner: the two- and 
three-particle phase space elements are, respectively
%----------------------------------------------------------------------
\vspace{-0.4cm}
%================================================

%t \\ \\ \\ \\ %\nl \nl \nl
%
%
%t \ea
%t \label{2-eeff}
%t\label{}\eqa
%
\vspace*{0.5cm}

\bqa
d\Phi_2(1,2) &=& \frac{d^3{\vec p}_1}{2p_1^0}
\frac{d^3{\vec p}_2}{2p_2^0}
\delta^4(p_{12}-p_1-p_2)\nonumber
\eqa

%======================================================================
and
%Three-particle phase space:
\bqa
d\Phi_3(1,2,3) &=& \frac{d^3{\vec p}_1}{2p_1^0}
\frac{d^3{\vec p}_2}{2p_2^0}\frac{d^3{\vec p}_3}{2p_3^0} 
\delta^4(p_{123}-p_{12}-p_3),\nonumber
\label{appx36}\eqa

which can be written as

\bqa
d\Phi_3(1,2,3) &=& 
d\Phi_2(1,2) \times ds' \times  d\Phi_2(12,3).\nonumber
\eqa
This connection is presented graphically in Figs. 5 and 6.
\vspace{-0.5cm}
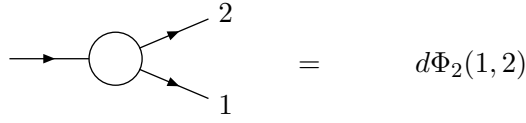
\begin{figure}[h]
\begin{center}
%                                               
%t\bqa
%t\ba{rcl}
%
%                                        12 to 1+2 
\begin{picture}(45,52)(0,25)
\ArrowLine(0,30)(30,30)
  \ArrowLine(49,34)(75,45)
  \ArrowLine(49,26)(75,15)
  \Text(79,44.0)[lb]{$2$}
  \Text(79,16.0)[lt]{$1$}
\CArc(40,30)(10,0,360)
\end{picture}
%t \hspace{2cm} & = & \hspace{1cm} 
   \hspace*{2cm}   =   \hspace*{1cm} 
$d\Phi_2(1,2)$
\vspace{0.5cm}
\caption{
\sf The 2-particle phase space. 
\label{2-f1}
}
\end{center}
\end{figure}

\vspace{-1.0cm}

%                                               
%\bqa
%\ba{rcl}
\begin{figure}[h]
\begin{center}
%
%                                        123 to 1+2+3 
\begin{picture}(45,52)(0,25)
\ArrowLine(0,30)(30,30)
\ArrowLine(50,30)(75,30)
  \ArrowLine(49,34)(75,45)
  \ArrowLine(49,26)(75,15)
  \Text(79,28)[lb]{$2$}
  \Text(79,44.0)[lb]{$3$}
  \Text(79,16.0)[lt]{$1$}
\CArc(40,30)(10,0,360)
\end{picture}
\hspace*{1.5cm}  =  \hspace{0cm} 
\begin{picture}(45,52)(0,25)
\ArrowLine(10,30)(30,30)
\CArc(40,30)(10,0,360)
  \Text(105,55)[lb]{$3$}
%  \ArrowLine(49,34)(75,45)
   \ArrowLine(49,34)(101,56)
\ArrowLine(50,30)(70,30)
%\ArrowLine(30,40)(30,50)
  \ArrowLine(89,34)(115,45)
  \ArrowLine(89,26)(115,15)
  \Text(119,44.0)[lb]{$2$}
  \Text(119,16.0)[lt]{$1$}
\CArc(80,30)(10,0,360)
\end{picture}
\hspace{3cm}  =  \hspace{0.3cm} 
$d\Phi_3(1,2,3)$
%\\ \\ \\ \\ %\nl \nl \nl
%
%
%\ea
%\label{2-eeff}
%\nl
%\label{}\eqa
%
\vspace*{0.5cm}
\caption{
\sf A sequential parametrization of the 3-particle phase space.
\label{2-f2}
}
\end{center}
\end{figure}
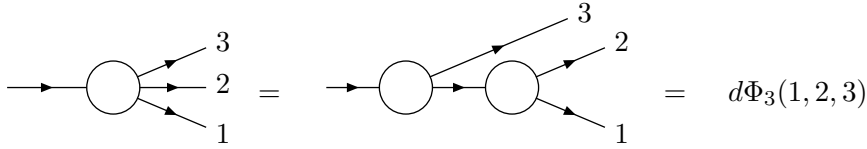
%----------------------------------------------------------------------
Thus the differential phase space volume $d{{\Phi}^{(3)}}$ 
finally is obtained as:\\

\bqa
d{\Phi_3}&=&(2\pi)^4\,\frac{d^3\vec{p_1}}{ (2\pi)^3 2 p^0_1}\,
\frac{d^3\vec{p_2}}{(2\pi)^3 2 p^0_2} 
\, \frac{d^3\vec{p}}{(2\pi)^3 2 p^0} 
\, \delta^4(\underbrace{k_1 + k_2}_{ = k_{12}}
\underbrace{-p_1-p_2}_{=-p_{12}}-p)
\nonumber\\
 &=& (2\pi)^{-5}\,d^4{p_1}\,
\delta(p_1^2-m^2) \, d^4{p_2}\,\delta(p_2^2-m^2)
\, d^4{p}\,\delta(p^2)\, 
\delta^4(k_{12}-p_{12}-p),
\nonumber
\eqa

which yields
\bqa
\label{dgamtot2}
%\rightarrow\quad 
d{\Phi_3{(3)}} &=& \frac{1}{(2\pi)^5}\,\frac{\pi}{16s}
\, d {\varphi_\gamma}\, d{V_2}\, d{s'}\, d {\cos\theta}
\nonumber\\
&=&
\label{dgamtot3}
\frac{1}{2}\frac{s}{(4\pi)^4}
\, d {\varphi_\gamma}
\, d\left(\frac{V_2}{s}\right)
\, d\left(\frac{s'}{s}\right)
\, d {\cos\theta} 
\nonumber\\
&=&
\label{dgamtot4}
\frac{1}{2}\frac{s}{(4\pi)^4}
\, d {\varphi_\gamma}\, d{x}\, d{r}\, d {\cos\theta},\nonumber
\eqa 
where the dimensionless variables
\bqa
\label{Rxvar}
r\equiv \frac{s'}{s}\qquad\mbox{and}\qquad x\equiv \frac{V_2}{s}
\nonumber
\eqa
have been introduced.
%\newpage
With the cross-section: \\
\vspace{-0.5cm}
\begin{eqnarray}
d\sigma(2\to 3) &=&\frac{(2 \pi)^4}{2 s }
 \times \left|{\cal M}\right|^2 \times~ d\Phi_3(3),\nonumber
\end{eqnarray}
%with the Kallen function:
%\bqa
%\lambda(a,b,c) &=& a^2 + b^2 + c^2 -2ab-2bc-2ca 
%\nonumber
%\eqa
%%%%%%%%%%%%%%%%%%%%%%%%%%%%%%%%%%%%%%%%%%%%%%%%%%%%
%Limits of integration 
%%%%%%%%%%%%%%%%%%%%%%%%%%%%%%%%%%%%%%%%%%%%%%%%%%%%
%
the integration can be performed analytically over the azimuthal
photon angle $\varphi_\gamma$:
\ba
\label{phigamma}
\varphi_\gamma\, \epsilon\, [0;2\pi].\nonumber
\ea
There are different reasons not to consider only the full
phase space but also parts of it. A real detector cannot see 
particles, which are too close to the beam pipe. This deficiency
can be taken into account by a constraint to $cos\theta$ 
(acceptance cut):
\ba
\label{ctcut}
-c \leq \cos\theta\leq c.\nonumber
\ea  
Calling the photon energy $E_{\gamma}$, there is a simple relation
between $E_{\gamma}$ and $r$, $r=1-2 E_{\gamma}/\sqrt{s}$. To keep
the photon energy larger than some cut $\omega$, i.e. 
$E_{\gamma} > \omega$, one has to constrain
\ba
\label{cut}
r \le 1 - 2 \omega/ \sqrt{s} = r_{\omega}. \nonumber
\ea
In some experiments, one wants to exclude events with hard photons,
i.e. one demands $E_{\gamma} \le E_{max}(\gamma)$. This cut translates
into the constraint
\ba
\label{con}
r = 1 - 2 E_{\gamma}/ \sqrt{s} \ge 1-2 E_{max}(\gamma)\sqrt{s}=r_{\gamma}.
\nonumber
\ea
In Fig.7 different acollinearity and photon energy cuts are shown.

\newpage

%
%--------------------------------------------------------------------------------
\begin{figure}[t]
\begin{flushleft}
\begin{tabular}{ll}
\hspace*{-1.5cm}
\mbox{
 \epsfig{file=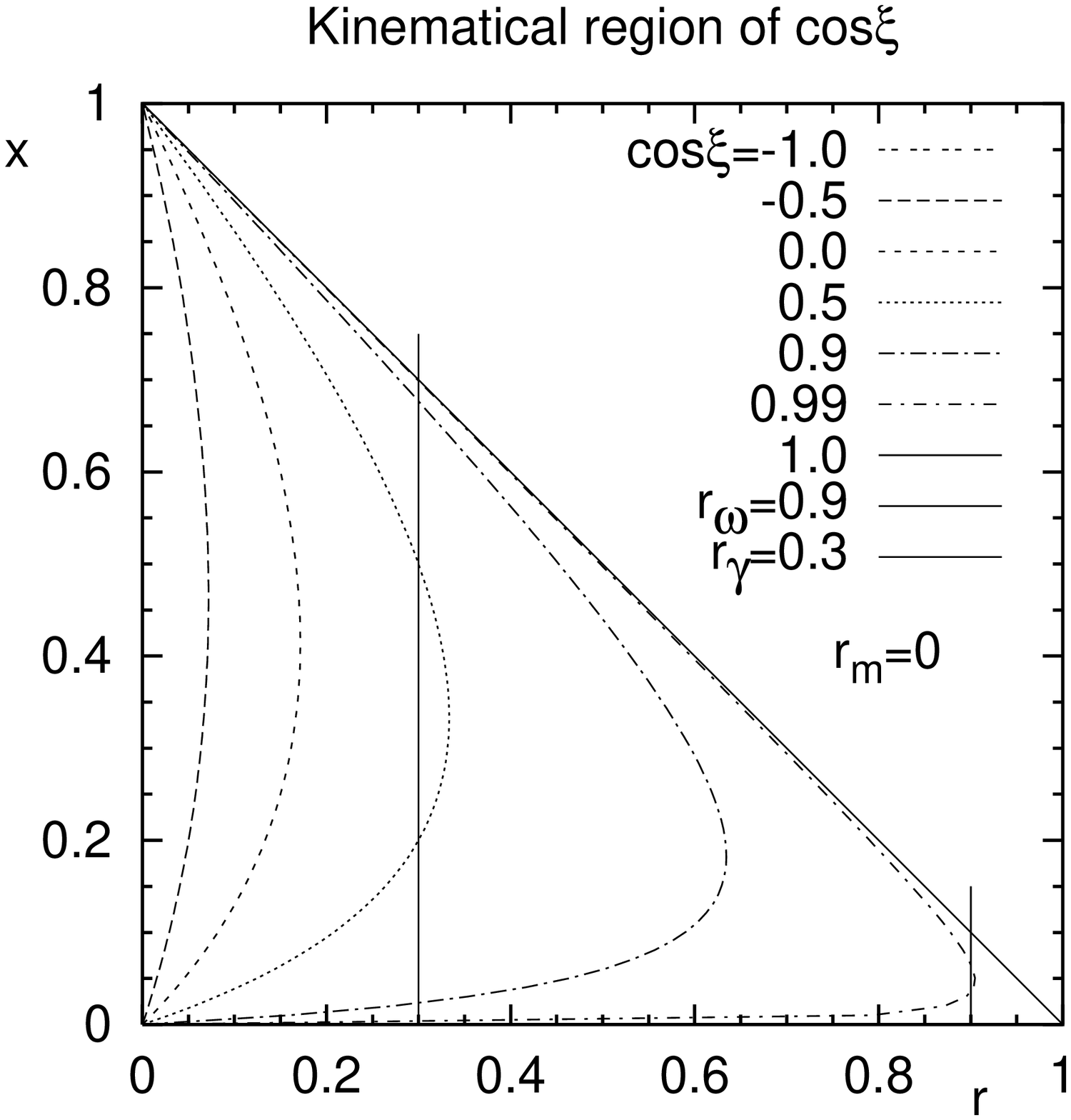,width=8.0cm}}
&
\hspace*{-1cm}
\mbox{
 \epsfig{file=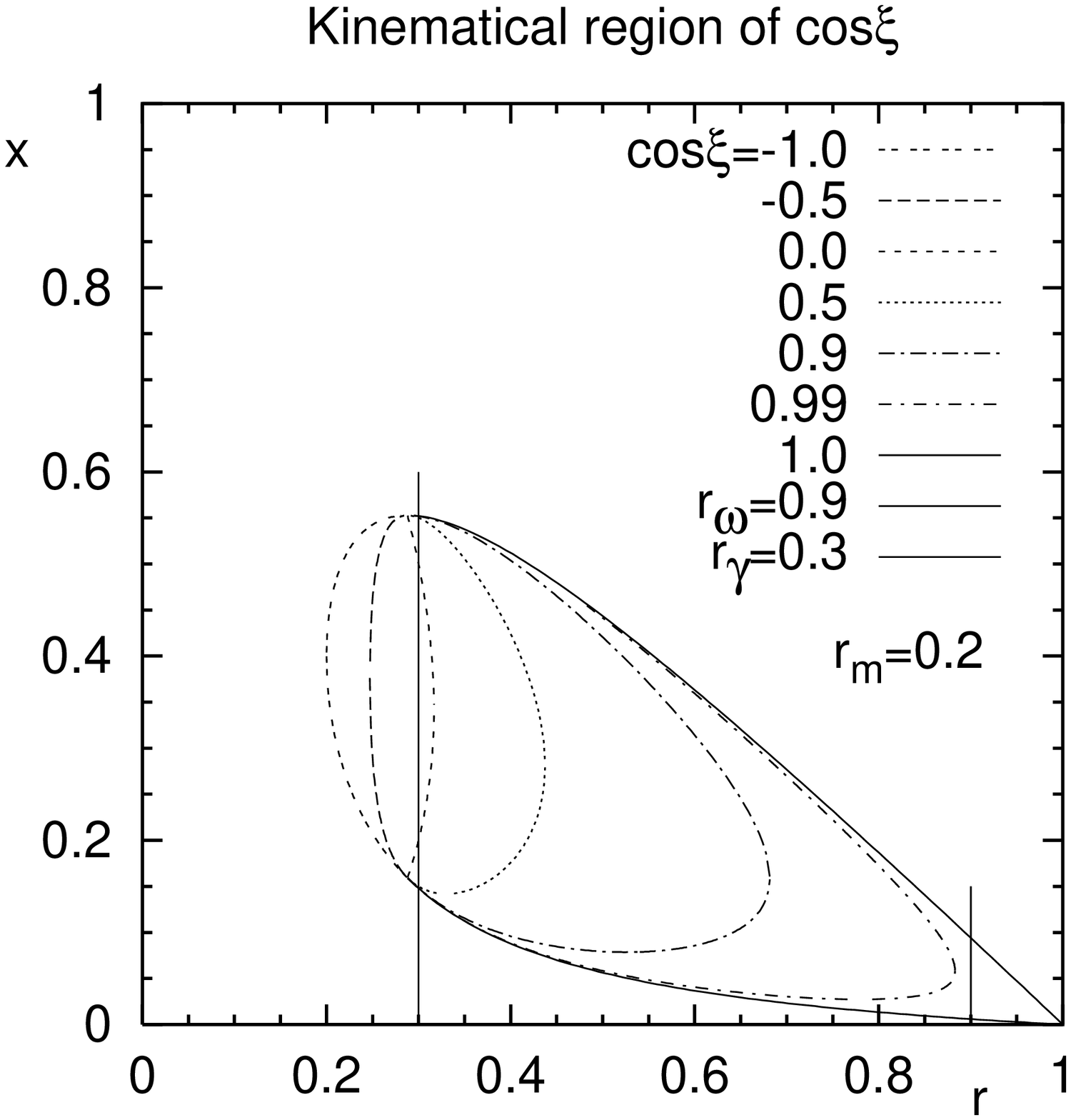,width=8.0cm}}
\end{tabular}
\end{flushleft}
\caption[Different cases in phase space with acollinearity cut]
{\sf
Phase space with acollinearity cut $R_\xi$: 
a. $r_m=\frac{4 m_t^2}{s}=0$ (zero top mass), 
b. $r_m=0.2$ (non-zero top mass);
$V_2/s\equiv x$.
\label{phasecuts}
}
\end{figure}
%------------------------------------------------------------------------------
%newpage
%---------------------------------------------------------------------

\vspace*{-1.0cm}

In the following we present the results of the {\tt topfit} and GRACE group
and their comparison. We present total cross sections ${\sigma}_\mathrm{tot}$
and forward~(F)-backward~(B) asymmetries, which in an obvious notation reads
%\begin{equation}
$A_{FB}=\frac{{\sigma}_{\mathrm{FB}}}{{\sigma}_{\mathrm{tot}}}$
%\end{equation}
with ${\sigma}_{\mathrm{tot}}={\sigma}_{F}+{\sigma}_{B}$ and 
     ${\sigma}_{\mathrm{FB}} ={\sigma}_{F}-{\sigma}_{B}$.
Further we separate QED and weak corrections: by QED we mean all 
IR-divergent diagrams (see Figs. 2,3 and boxes)
including their counterterms plus born~(0) and soft
contributions. The Standard Model (SM) means QED plus weak, where 
the fermion contributions in the photon selfenergy
and the charge renormalization are counted as `weak'.
The notion '$\mathrm{tot}$' also means inclusion of hard photons
(integrated over the whole phase space).\\

\input{1table}$
$
%newpage
%breakpage
\include{2tables_japan12032002}

\include{plots}

In Table 1 the results are given for total cross sections for two
energies and in Tables 2,3 for differential cross sections. 
If not stated otherwise, the parameter input is the same
as in Ref. \cite{Fleischer:2002rn}.  We see that
the agreement is indeed excellent. As a general statement
we can say, that the accuracy of the GRACE numbers for the 
non-hard-cross-sections is higher than shown while for the hard 
cross sections the same applies for the {\tt topfit} numbers.

   Concerning the `elastic' entities in Table 1
we have in general an agreement of up to six decimals while in the
hard ones the maximum error is contained in $A_{FB}$, which is 
2 $\times 10^{-4}$ (.58 per mille). Concerning Table 2 we see that the
differential Born cross sections agree up to 12 decimals. We also
show the QED contributions for $\omega/\sqrt s=0.1$ since they are
slightly more precise than those shown in Ref. \cite{Fleischer:2002rn}
due to a higher accuracy of the soft interference term. For
practical applications this large soft photon cutoff is, however,
not applicable. The agreement in Table 1
is least good for those results which include hard
bremsstrahlung, but we have for $\omega/\sqrt s=0.00001$ an 
agreement of 3 decimals. Presumably the errors given by the GRACE MC
calculations are somewhat underestimated. Nevertheless the
results of the GRACE group for  
${\sigma}_{\mathrm{tot}}$ and $A_{\mathrm{FB}}$
in Table 1 are more precise than those for the differential cross
sections in Table 2.

   In Figs. 8 and 9 we present our results for the differential
cross sections for $e^+e^- \to c {\bar c}$ and $e^+e^- \to t {\bar t}$
and the corresponding percentage corrections.
For the $c {\bar c}$ channel we observe huge corrections
in particular in forward and backward direction. These corrections
are reduced drastically by cutting hard photons  and are therefore 
not of real physical relevance. For $t {\bar t}$ production the mass
serves as a cutoff.

{\bf Ackknowledgement:} One of us (J.F.) wants to thank DESY for
extended kind hospitality.

\small
\bibliographystyle{utphys_}
\bibliography{toppair}

\end{document}

%% file: 1table.tex
\begin{table}[H]
$$
\begin{array}{|r|c|c|c|c|c|c|}
\hline 
\vrule height 3ex depth 0ex width 0ex
 \sqrt{s}~~    &\sigma_{\mathrm{tot}}^0 & A_{\mathrm{FB}}^0& \sigma_{\mathrm{SM,tot}}  &\sigma_{\mathrm{SM,FB}}  & \sigma_\mathrm{tot} & A_{\mathrm{FB}}
\\ 
\hline 
\hline  
500 & \mathrm{T:}~ 0.5122744 & 0.4146039 & -0.1198972 & -0.0855551 &0.526337  &  0.362929
\\ 
    & \mathrm{G:}~ 0.5122751 & 0.4146042 & -0.1198973 &           & 0.526371 & 0.363140
\\
\hline
1000 &\mathrm{T:}~  0.1559185 & 0.5641706 & -0.0683693 & -0.0522582 &
0.171916& 0.488869
\\
   & \mathrm{G:}~  0.1559187 & 0.5641710 & -0.0683695 &            & 0.171931 & 0.488872
\\ 
\hline
\end{array}
$$
\caption{
\sf Total cross sections (in pbarn) and forward-backward asymmetries.
${\sigma}^0_\mathrm{tot}$ (Born) and ${\sigma}_\mathrm{SM,tot}$ are `elastic'
and ${\sigma}_\mathrm{tot}$ includes hard photons, $\omega/\sqrt{s}$ = 0.00001.
%
%Integrated elastic and hard cross-section in pbarn  and integrated
%cross-section and forward-backward asymmetry; no cuts applied,
}
\end{table}

%\end{document}

%% file: 2tables_japan12032002.tex
%\documentclass[12pt]{article}
%\usepackage{amsmath,a4wide}

%\begin{document}

% Born         : iborn=1 isoft=0 ivirt=0 iweak=0 iqed=0 iqedaa=0 iphotm=0
% Born+QED     : iborn=1 isoft=1 ivirt=1 iweak=0 iqed=1 iqedaa=0 iphotm=0
% Born    +weak: iborn=1 isoft=0 ivirt=1 iweak=1 iqed=0 iqedaa=1 iphotm=0
% Born+QED+weak: iborn=1 isoft=1 ivirt=1 iweak=1 iqed=1 iqedaa=1 iphotm=0
%
% SM (el.)     : iborn=1 isoft=1 ivirt=1 iweak=1 iqed=1 iqedaa=1 iphotm=0
% SM+hard      : iborn=1 isoft=1 ivirt=1 iweak=1 iqed=1 iqedaa=1 iphotm=0
%                ihard=1

%\section{Numerical comparison of cross-sections}
% 11 March 2002 best numbers available used
% Japan grace: 27 Feb 2002, Subject of mail from Fujimoto: our results
\vspace*{5mm}
%############################################################################
%############################################################################
%###################### 500 #################################################
%############################################################################
%############################################################################
\begin{table}[H]
%$\sqrt s = 500$ GeV:
$$
\begin{array}{|r|l|l|l|l|l|}
\hline 
\vrule height 3ex depth 0ex width 0ex
\cos\theta & ~~~\omega / \sqrt{s} & \sig_{\mathrm{Born}} & \sig_{\mathrm{QED}} & \sig_{\mathrm{SM}}
&  \sig_{\mathrm{tot}}
\\ 
\hline 
\hline  
-0.9 & \mathrm{T:}~0.1 
&  0.108839194075  & +0.098664253  & +0.11408410 & 0.13144
\\
     & \mathrm{T:}~0.00001 
& 0.108839194075   & -0.017474702  & -0.002054858 & 0.13229 
\\[-0.2mm]
     & \mathrm{G:}~0.00001
&  0.108839194076  &               & -0.002054859  & 0.13206(12)
\\
\hline 
-0.5 & \mathrm{T:}~ 0.1     
&  0.142275069392  &  +0.12850790 & +0.14308121  & 0.15973
\\
     & \mathrm{T:}~ 0.00001 
&   0.142275069392 &  -0.029702340 & -0.015129038 & 0.16029
\\
     & \mathrm{G:}~0.00001
&  0.142275069393  &               & -0.015129039    & 0.16013(13)
\\
\hline 
+0.0&  \mathrm{T:}~0.1    
&  0.225470464033  & +0.20239167  & +0.21718801  & 0.23638
\\
     & \mathrm{T:}~0.00001
&   0.225470464033 &  -0.058010508 & -0.043214169 & 0.23476
\\
 &  \mathrm{G:}~0.00001    
&  0.225470464033  &               & -0.043214168   & 0.23513(14) 
\\
\hline 
+0.5 &  \mathrm{T:}~0.1    
&  0.354666470332  & +0.31511723  & +0.32933727  & 0.35651
\\
    & \mathrm{T:}~0.00001
&   0.354666470332 & -0.109721291  & -0.095501257 & 0.35062
\\
 & \mathrm{G:}~0.00001   
&  0.354666470332  &               & -0.095501252 & 0.35104(17)
\\\hline 
+0.9 & \mathrm{T:}~0.1     
&  0.491143715767  &  +0.43071437 & +0.44290816  & 0.48796
\\
      & \mathrm{T:}~0.00001
&  0.491143715767  &  -0.179672655 & -0.16747886  & 0.47768
\\
 & \mathrm{G:}~0.00001  
&  0.491143715767  &               &  -0.16747886  & 0.47709(21)
\\ \hline
\end{array}
$$

\caption{\sf Various differential cross sections (see also text).
  The upper and lower numbers correspond to the {\tt topfit} (T) 
  and GRACE (G)
  approach, respectively, $\sqrt{s}$ = 500 GeV.
%The accuracy of the GRACE numbers for the non-hard cross-sections is
%higher than shown, for the hard corrections the same applies for the
%topfit numbers.
}
\end{table}

%############################################################################
%############################################################################
%############################################################################
%############################################################################
%#####################  new 1000  ###########################################
%############################################################################
%############################################################################
\begin{table}[H]
%$\sqrt s = 1000$ GeV:
$$
\begin{array}{|r|l|l|l|l|l|}
\hline 
\vrule height 3ex depth 0ex width 0ex
\cos\theta & ~~~\omega / \sqrt{s} & \sig_{\mathrm{Born}} & \sig_{\mathrm{QED}} & \sig_{\mathrm{SM}}
&  \sig_{\mathrm{tot}}
\\ 
\hline 
\hline  
-0.9 & \mathrm{T:}~0.1 
  &  0.0227854232732  &  +0.020365844  & +0.023101706  & 0.036334
\\
     & \mathrm{T:}~0.00001 
  &     0.0227854232732 &  -0.004756230   & -0.002020367 & 0.036461
\\[-0.2mm]
     & \mathrm{G:}~0.00001
  &  0.02278542327319 &                &  -0.002020369  & 0.036582(48)
\\
\hline 
-0.5 & \mathrm{T:}~0.1     
  &  0.0297821311031  &  +0.026741663  & +0.028823021  & 0.038888
\\
     & \mathrm{T:}~0.00001 
     &  0.0297821311031 &  -0.008561495 & -0.006480137  & 0.039055
\\
     &\mathrm{G:}~0.00001
     & 0.0297821311031  &               & -0.006480139  &0.038965(42)
\\
\hline 
+0.0& \mathrm{T:}~0.1    
&  0.0611800674224  &  +0.054539344  & +0.054950889  & 0.067789
\\
     &\mathrm{T:}~0.00001
&  0.0611800674224  &  -0.021532420    &-0.021120874 &  0.067801
\\
     & \mathrm{G:}~0.00001    
&  0.0611800674225  &                &  -0.021120874 & 0.068039(55)
\\
\hline 
+0.5 & \mathrm{T:}~0.1    
&  0.117746949888   &  +0.10311626   & +0.099416999   & 0.12095
\\
    &\mathrm{T:}~0.00001
&  0.117746949888   &  -0.050123708  & -0.053822973   &  0.12051
\\
 &\mathrm{G:}~0.00001   
&  0.117746949888   &                & -0.053822964    & 0.12064(07)
\\
\hline 
+0.9 &\mathrm{T:}~0.1     
&  0.181122097086   &  +0.15403823   & +0.14426232  & 0.19355
\\
      &\mathrm{T:}~0.00001
&  0.181122097086    & -0.096682759 &  -0.10645866 & 0.19272
\\
    &\mathrm{G:}~0.00001  
&  0.181122097086   &                &  -0.10645866   & 0.19057(10)
\\ 
\hline
\end{array}
$$
%\bigskip

\caption{\sf
Same as Table 2, $\sqrt{s}=1000$ GeV.
}
\end{table}
%############################################################################
%############################################################################
%############################################################################

%\end{document}

%% file: plots.tex
\newpage

\begin{figure}[H]
\hspace{-1cm}
\begin{minipage}[l]{9cm}
 \centerline{\epsfysize=9cm\epsffile{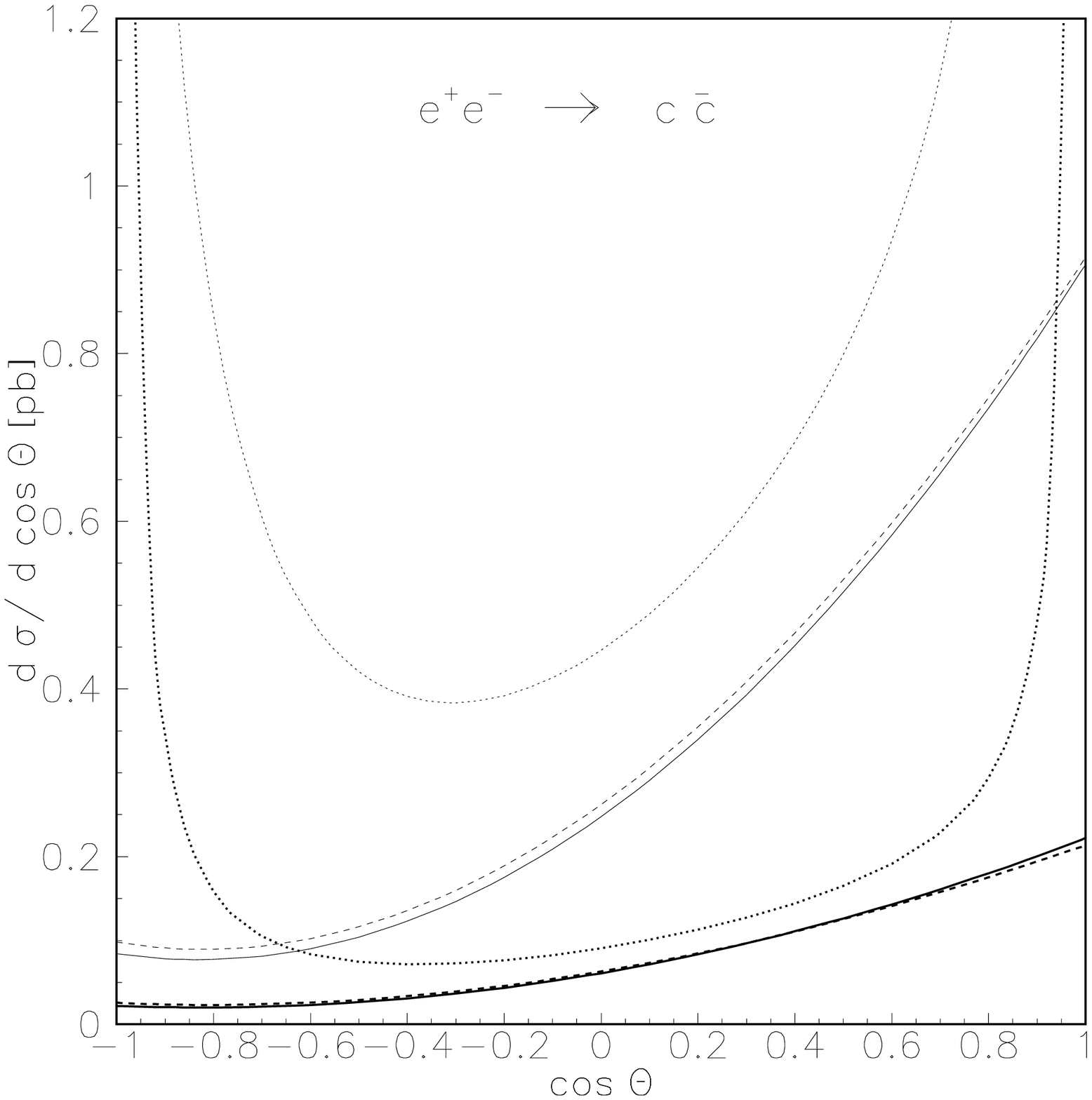}}
\vspace{-1cm}
\end{minipage}
\begin{minipage}[c]{1cm}
$ $
\end{minipage}
\begin{minipage}[r]{9cm}
%\vspace{-1.9cm}
\hspace{-1cm}
 \centerline{\epsfysize=9cm\epsffile{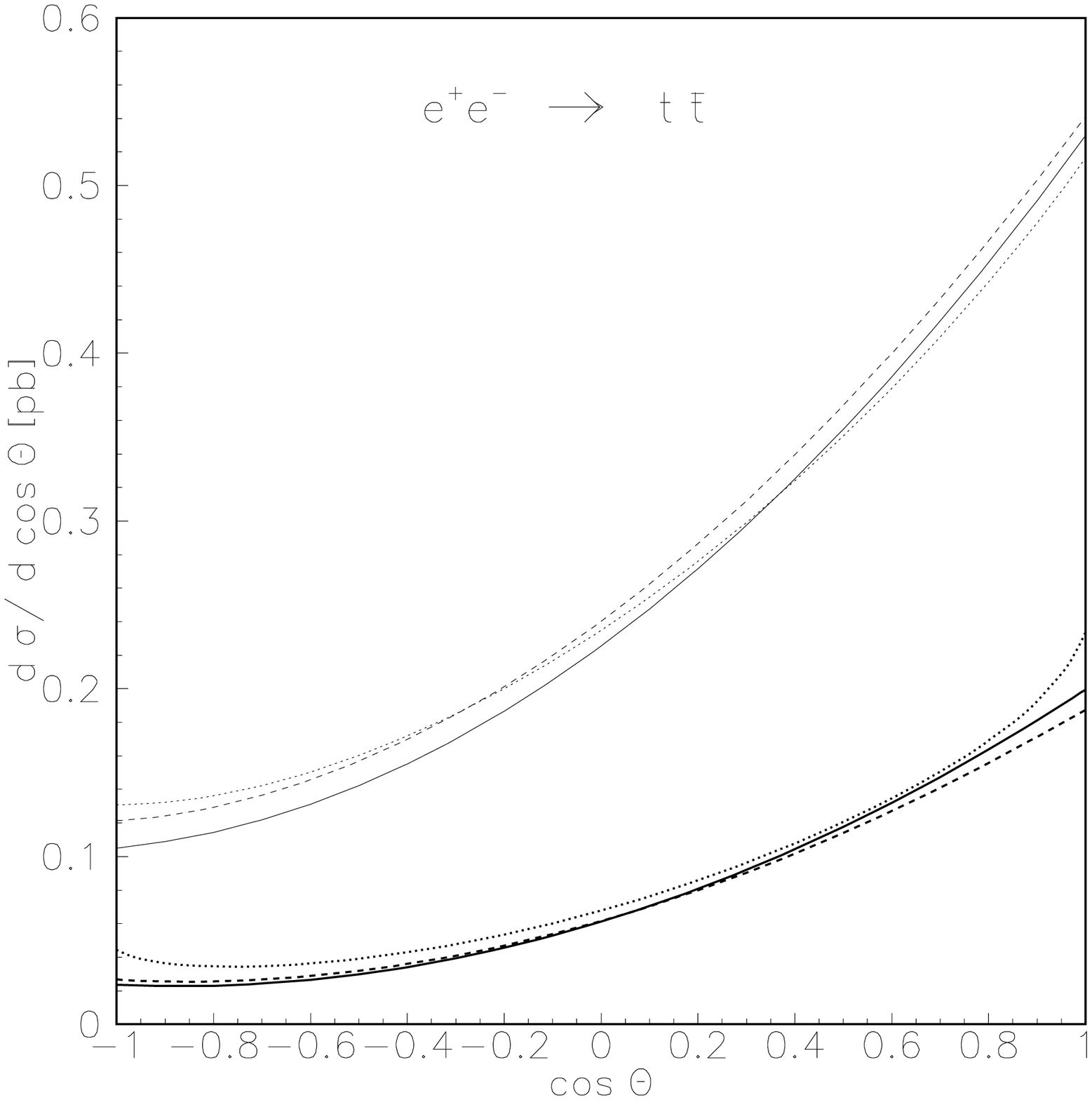}}
\vspace{-1cm}
%\caption{\label{eett}{Same as Fig. (\ref{eecc}) for the process
% $e^+e^-$ $ \to$ $ t \bar{t}$.}}
\end{minipage}
\caption{\label{eecc}{\sf Differential hard photon cross sections 
${\frac{d \sigma}{d cos \theta}}$
 of the processes $e^+e^- \to c \bar{c}$ and $e^+e^- \to t \bar{t}$. 
Solid, dashed and dotted lines stand for
Born, weak and SM, respectively. Thin lines correspond to
$\sqrt{s} = 500 $~GeV and thick lines to $\sqrt{s} =1000 $~GeV.}}
\end{figure}

\vspace*{0.2cm}
\hspace{-1cm}

\begin{figure}[H]
\hspace{-1cm}
\begin{minipage}[l]{9cm}
 \centerline{\epsfysize=9cm\epsffile{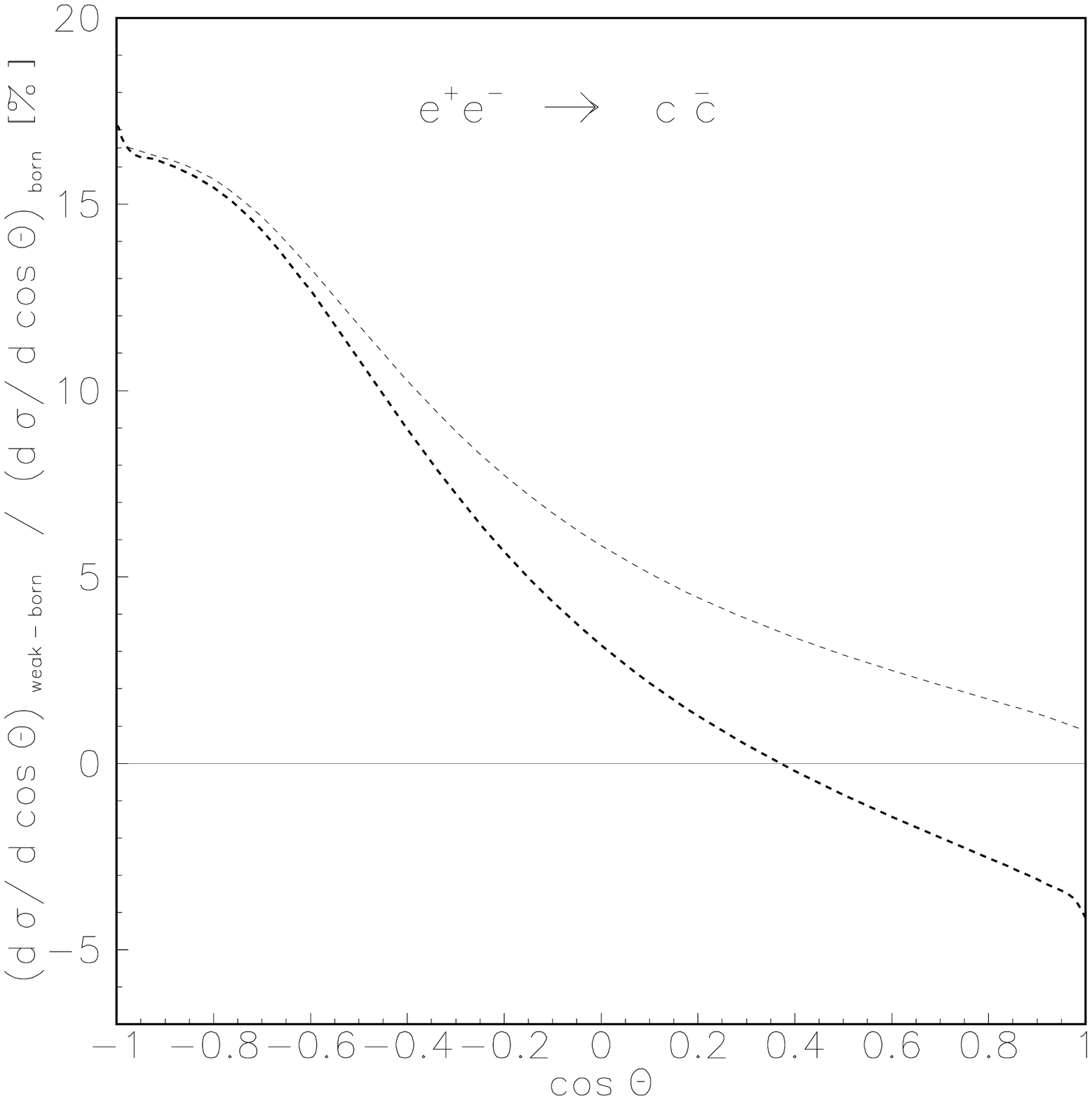}}
\vspace{-1cm}
\end{minipage}
\begin{minipage}[c]{1cm}
$ $
\end{minipage}
\begin{minipage}[r]{9cm}
\hspace{-1cm}
 \centerline{\epsfysize=9cm\epsffile{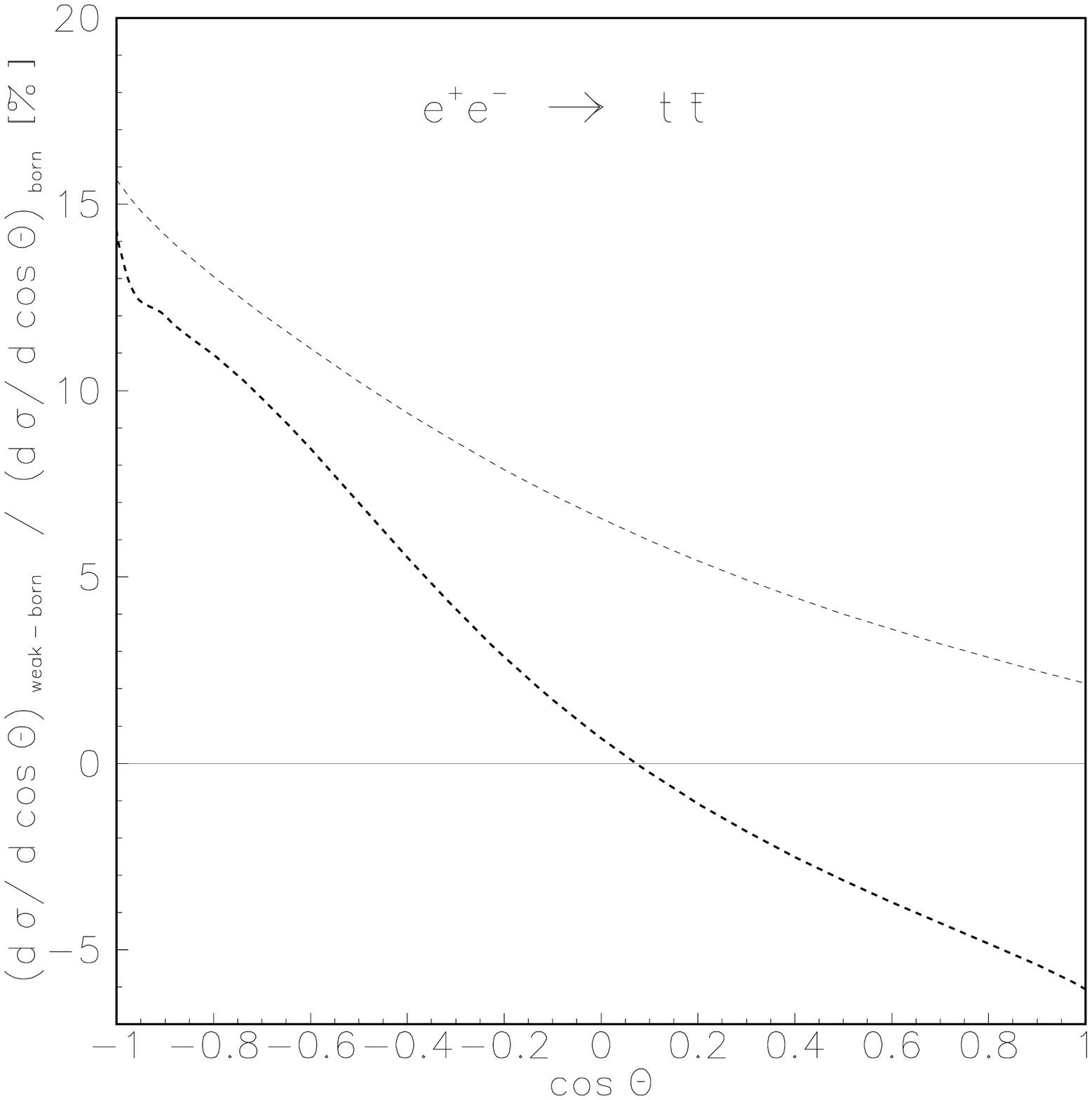}}
\vspace{-1cm}
%\caption{\label{eettrel}{Same as Fig. (\ref{eeccrel}) for the
%process
% $e^+e^- $ $\to$ $ t \bar{t}$.}}
\end{minipage}
\caption{\label{eeccrel}{\sf One-loop percentage corrections
with respect to Born, corresponding to Fig.8.
Thin lines stand for
$\sqrt{s} = 500 $~GeV and thick lines for $\sqrt{s} =1000 $~GeV.}}
\end{figure}